\def\year{2016}
\documentclass[letterpaper]{article}
\usepackage{./styles/aaai}
\usepackage{times}
\usepackage{helvet}
\usepackage{courier}
\usepackage{amsmath}
\usepackage{amsfonts}
\usepackage{amssymb}
\usepackage{array}
\usepackage{graphicx}
\usepackage{setspace}
\usepackage{mathptmx}
\usepackage{url}

\newcommand{\makefigure}[3]{
  \begin{figure}[#1]
  \begin{center}
  \includegraphics[width=84mm]{./images/#2}
  \end{center}
  \vspace{-1.5mm}
  \caption{#3}
  \label{fig:#2}
  \end{figure}
}

\newcommand{\furl}[1]{\footnote{\url{#1}}}

\newcolumntype{x}[1]{>{\centering\hspace{0pt}}p{#1}}

\newcommand{\tn}{\tabularnewline\hline}

\newcommand{\caplab}[2]{\caption{#1}\label{#2}}

\newcommand{\conftbl}{\centering\footnotesize\vspace{5pt}}

\begin{document}

\title{GECKA3D:\\A 3D Game Engine for Commonsense Knowledge Acquisition}

\author{
Erik Cambria\\
School of Computer Engineering\\
Nanyang Technological University\\
cambria@ntu.edu.sg
\And
Tam V. Nguyen \\
ARTIC Research Center\\
Singapore Polytechnic\\
nguyen\_van\_tam@sp.edu.sg
\And
Brian Cheng\\
ARTIC Research Center\\
Singapore Polytechnic\\
brian\_cheng@sp.edu.sg
\AND
Kenneth Kwok\\
Institute of High Performance Computing\\
Agency for Science, Technology and Research\\
kenkwok@ihpc.a-star.edu.sg
\And
Jose Sepulveda\\
ARTIC Research Center\\
Singapore Polytechnic\\
sepulveda\_jose@sp.edu.sg
}
\pdfinfo{ 
/Title (Semantic Outlier Detection) 
/Author (Erik Cambria, Giuseppe Melfi) 
}

\maketitle
\begin{abstract}
Commonsense knowledge representation and reasoning is key for tasks such as artificial intelligence and natural language understanding. Since commonsense consists of information that humans take for granted, gathering it is an extremely difficult task. In this paper, we introduce a novel 3D game engine for commonsense knowledge acquisition (GECKA3D) which aims to collect commonsense from game designers through the development of serious games. GECKA3D integrates the potential of serious games and games with a purpose. This provides a platform for the acquisition of re-usable and multi-purpose knowledge, and also enables the development of games that can provide entertainment value and teach players something meaningful about the actual world they live in.
\end{abstract}

\section{Introduction}
Many existing multimedia annotation and artificial intelligence  systems benefit from large-scale datasets with accurate annotations. For example, $40$ participants were gathered to rate the attractiveness scores for $1240$ subjects in Beauty Sense project~\cite{BeautySense,TOMM}. In another work, $69$ participants  were requested to provide touch saliency for at least $400$ images~\cite{Mengdi}. However, constructing such datasets is generally tedious and labor-intensive, as it involves massive manual-cropping and hand-labeling. Therefore, games with a purpose (GWAPs) are a simple yet powerful means to collect useful information from players in a way that is entertaining for them. A GWAP embeds  data collection implicitly into the game playing process. It is developed from the human computation technique which relies on the collaboration of game players to solve problems. Each player performs a small part of a massive computation, similar to a processor in a distributed system. Unlike computer processors, however, humans require a certain incentive to participate in such collective computation. GWAP aims to implicitly embedding the computational tasks in an enjoyable and attractive games while guaranteeing correct solutions to the problems. 

Over the past few years, many GWAPs~\cite{ahnlab,ahngam,labelme,phog} have been proposed to exploit the brainpower made available by multitudes of casual gamers to perform tasks that, despite being relatively easy for humans to complete, are difficult for machines. GWAPs are best known for image annotation. In the `ESP' game \cite{ahnlab}, players guess content objects or properties of random images by typing what they see when it appears on the screen. Other image annotation games include: Matchin \cite{hacmat}, a two-player game in which each player is shown two images and asked to click on the image their partner would prefer, and Phetch \cite{ahnimp}, a game that collects explanatory descriptions of images in order to improve Web accessibility for the visually impaired.  Recently,  Purposive Hidden-Object-Game~\cite{phog} was proposed. It was designed to imperceptibly embed localizing objects into an enjoyable game process thus attracting many people to make voluntary contribution to annotating images.

Besides images, GWAPs have been used for video annotation. For example, OntoTube \cite{sioont}, Yahoo's Videotaggame \cite{zwovid}, and Waisd \cite{add100}, are all games in which two players have to quickly agree on a set of tags for the same streaming YouTube video. GWAPs have also been exploited to automatically tag music tracks with semantic labels. HerdIt \cite{baruse}, for example, asks players to accomplish various tasks and answer quizzes related to the song they are listening to. In Tagatune \cite{lawtag}, two players listen to an audio file and describe to the other what each is hearing. Players must then decide whether or not the game has played the same soundtrack to both participants. 

One of the most interesting tasks GWAPs can be used for is commonsense knowledge acquisition from members of the general public \cite{chklea,speope,camacq}. One example, Verbosity \cite{ahnver}, is a real time quiz game for collecting commonsense facts. In the game, two players take different roles at different times: one functions as a narrator, who has to describe a word using templates, while the other has to guess the word in the shortest time possible. 
Another example is FACTory Game \cite{lencyc}, a GWAP developed by Cycorp which randomly chooses some facts and presents them to players in order for them to guess whether a statement is true or false. A variant of the FACTory game is the Concept Game on Facebook \cite{herthe}, which collects commonsense knowledge by proposing random assertions to users and asks them whether the given assertion is meaningful or not. 

Questions are very important in collecting commonsense. Kuo and Hsu~\cite{Kuo} introduced the problem of resource-bounded crowdsourcing of commonsense knowledge and proposed an approach to find a productive question set automatically. Another problem with existing GWAPs is that the information gathered from them is often unrecyclable; acquired data is often applicable only to the specific stimuli encountered during gameplay.  Moreover, such games often have a fairly low `sticky factor', and are often unable to engage gamers for more than a couple of minutes. The fun factor sometimes is not considered during the design of such a GWAP system.

In this paper, we propose a 3D game engine for commonsense knowledge acquisition (GECKA3D), that aims to overcome the main drawbacks of traditional data-collecting games by empowering users to create their own GWAPs and by mining knowledge that is highly reusable and multi-purpose. In addition, we also consider adding game mechanism to enhance the fun factor to engage the players. In particular, the GECKA3D framework has two modes, editor and player. The editor mode allows game designers to create compelling serious games for their peers to play. In doing so, they gather commonsense knowledge useful for intelligent applications in any field requiring in-depth knowledge of the real world, including reasoning, perception and social systems simulation \cite{camcomshort}. Meanwhile, the player mode is used by players to perform the task. The framework was built on top of the cocos2d game engine\footnote{http://www.cocos2d-x.org}. Note that the cocos2d game engine is a suite of open-source, cross-platform, and game-development tools used by thousands of developers all over the world.

\section{Proposed Framework}
An important difference between traditional artificial intelligence systems and human intelligence is the human ability to harness commonsense knowledge gathered from a lifetime of learning and experience to make informed decisions~\cite{camacq}. This allows humans to adapt easily to novel situations where artificial intelligence fails catastrophically due to a lack of situation-specific rules and generalization capabilities. Commonsense knowledge also provides background information enabling humans to successfully operate in social situations where such knowledge is typically assumed \cite{camaf2}. To this end, GECKA3D allows users to design compelling serious games via editor mode. As opposed to traditional GWAPs, GECKA3D does not limit users to specific, often tedious, tasks, but rather gives them the freedom to choose both the kind and the granularity of knowledge to be encoded, through a user-friendly and intuitive interface. Not just a system for the creation of microgames, GECKA3D is a serious game engine that aims to give designers the means to create long adventure games to be played by others. GECKA3D offers functionalities typical of role-play games (RPGs), e.g., a question/answer dialogue box enabling communication and the exchange of objects (optionally tied to correct answers) between players and virtual world inhabitants, a library for enriching scenes with useful and yet visually-appealing objects, backgrounds, characters.

\makefigure{t}{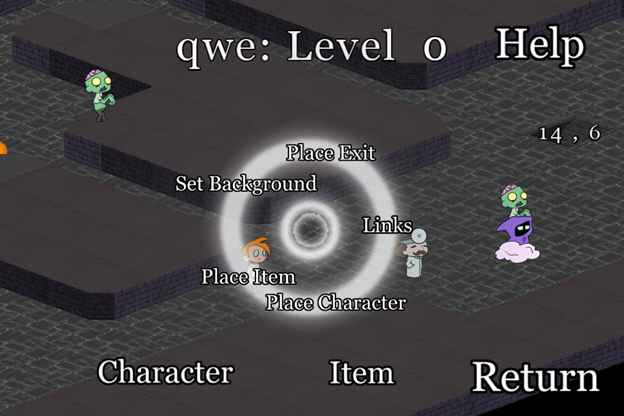}{Editor mode. Game designers can drag and drop objects and characters from the library and specify how these interact with each other.}

\subsection{Initial Knowledge Construction}
The core elements in the proposed GECKA3D framework are objects and actions. Their semantics can be specified through prerequisite-object-action-goal (POAG) quartets. Prerequisites indicate what must be present or have been done before applying the object or action. Objects indicate states or objects of the world (including emotional states, e.g., ``if I give food to hungry people, their happiness is likely to rise''). Goals in turn specify the specific scene goals that are facilitated by that particular POAG quartet. In order to provide designers with initial knowledge, we first build the corpus of Object-Action groups. We consider verbs as actions whereas nouns can be regarded as objects. In the initial step of corpus retrieval, we utilized the combinations of verbs and nouns as the search terms fed into Microsoft Bing search engine\footnote{http://www.bing.com} in the form of `how to + a $<$noun$>$'. The number of searched results can be considered as the joint likelihood of the combination of action and object. We also integrate boost library\footnote{http://www.boost.org} for semantic parsing. In the post-processing step, we completed cleaning of verb and noun corpuses with normalization, e.g., forcing all characters to lowercase, and removing html tags. As a result, we collect the initial corpus with 1500 nouns and 636 verbs in English.

\subsection{Editor Mode}
Unlike other serious games or data collection tools~\cite{ahnpee,labelme,phog,TamMM13}, GECKA3D provides authoring tools (or so called editor mode) for designers to create the training scenarios. The editor mode is built in 2.5D which is similar as 2D graphical projections and simulates the appearance of being three-dimensional (3D). In other words, the gameplay is in a 3D video game that is restricted to a 2D plane or has a virtual camera with a fixed angle. 

This is inspired by games such as SimCity\footnote{https://en.wikipedia.org/wiki/SimCity}. Figure~\ref{fig:gecka3dscene1}  illustrates the editor mode in being used by a designer. 
The designers are able to use the editor to create the training scenario in the form of tile based levels. The GECKA3D framework allows designers to define actions, objects, and goals. In case an action or an object is not available in the library, GECKA3D allows game designers to define their own custom items by building shapes from a set of predefined geometric forms or applying transformations to existing items. This enables the creation of new objects for which there is no available icon by combining available graphics and predefined shapes, and the use of transformations to create various object states, such as a `broken cup'. The ability of users to create their own custom items and actions is key to maintaining an undisrupted game flow. 

Each game has enter/exit portal sprites. The designer has to define the starting location of the in-game character. It is easy for designer to know which portal when defining entrance. This feature allows the character to traverse in different levels. As mentioned in~\cite{ahngam}, people play games not because they are personally interested in solving an instance of a computational problem but because they wish to be entertained. Therefore, in order to gain the sticky factor of the game, GECKA3D embeds some game mechanism. For example, it allows designers to add monsters/zombies to chase the in-game player. Therefore, the player has to make decision quickly in order to avoid the pursuers. 

\makefigure{!t}{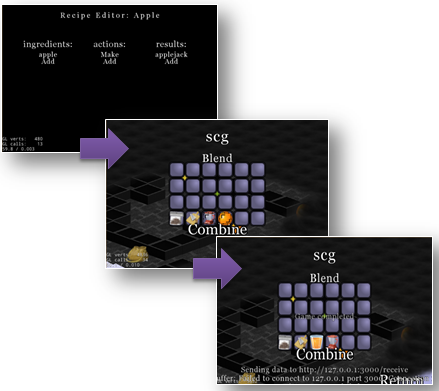}{The designers specify the item combination rule. They also define the outcome of the combination. Note that items can be objects or actions.}

\subsection{Object-Action Combinations}
Game designers drag and drop objects and characters from action/object libraries into scenes. In addition, the designers can add item combinations and the corresponding outcomes, for example, orange + blender = orange juice. In particular, designers can create, load and delete such item combinations. They can specify POAG quartet by grouping objects with the actions performed over it (Fig.~\ref{fig:gecka3dprerequisites}). POAG quartets give us pieces of commonsense information like ``the result of blending an orange with a blender, is orange juice''.

Towards the goal of improving gameplay, and since we are mainly interested in typical commonsense knowledge, POAG quartets associated with a specific object type are shared among all the instances of such an object (`inheritance'). Whenever a game designer associates a POAG to an object in the scene, that POAG instantly becomes shared among all the other objects of the same type, no matter if these are located in different scenes. New instances inherit this POAG as well. Game designers, however, can create exceptions of any object type through the creation of new custom objects. A `moldy bread' custom object, for example, normally inherits all the POAGs of `bread' but these can be changed, modified, or removed at the time of object instantiation without affecting other `bread' type objects.

The POAG specification is among the most effective means to collect commonsense knowledge, given that it is performed quite often by the game designer during the creation of scenes.
From a simple POAG definition we may obtain a large amount of knowledge, including interaction semantics between different objects, prerequisites of actions, and the goals commonly associated with such actions. These pieces of commonsense knowledge, are very clearly-structured, and thus easy to assimilate into the knowledge base, due to the fixed framework for defining interaction semantics.

\makefigure{!t}{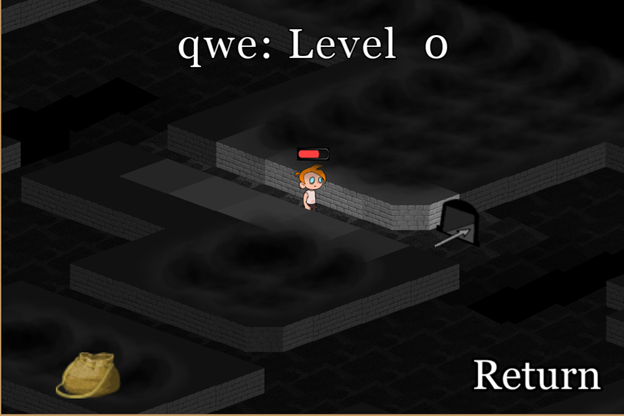}{Status of an in-game character in the scene. The in-game character has to fulfill the goal of the game, meanwhile he has to avoid the attack of the embedded zombies. }
\subsection{Player Mode}
Besides allowing for the acquisition of knowledge from game designers, GECKA3D enables players of the finished games to be educated in useful ways, all while being entertained. 

\makefigure{h}{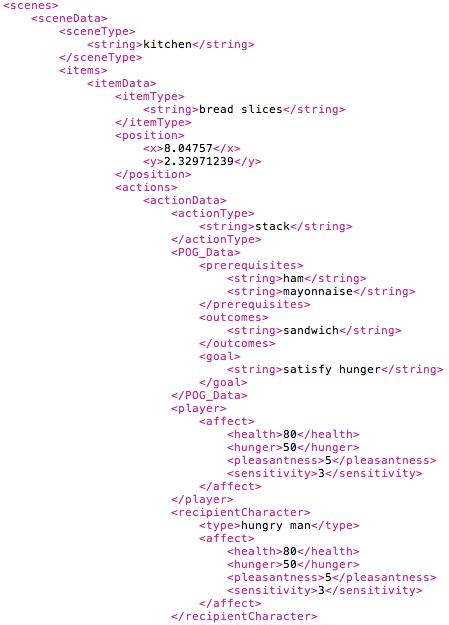}{A sample XML output deriving from the creation of a scene in GECKA3D. Actions are collected and encoded according to their semantics.} 

In particular, the players use the player mode to control the in-game character to fulfill the predefined goals. The gameplay is also built in 2.5D as similar as the editor mode. As aforementioned, the player has to make decisions quickly to avoid the zombies or monsters. The in-game rooms and corridors are randomly generated to make sure the players have no bias about the spatial layout. Inspired by other games such as Diablo\footnote{https://en.wikipedia.org/wiki/Diablo\_(video\_game)}, only a small portion of view close to in-game character is revealed. The more the character moves, the clearer vision the player achieves.  We also utilize the path finding algorithm for player movement and explorer. When the user clicks to one point in the map, the system automatically finds the shortest path from the user's current position and the target point. The in-game zombies are inserted to put pressure on the player. Note that the zombies chase the player in a turn based manner. 

\subsection{Server}
Server is yet another important component of the GECKA3D framework. It is used to store all of the information of the gameplay. 
The core functionality include game data uploading from remote clients to server. Note that our editor mode and player mode are deployed in many platforms thanks to the cross-platform characteristic of the chosen game engine. In particular, we use mongoDB and nodejs for server behavior. Http requests are integrated into mobile communication. Data can be transferred from phone to arbitrary server on internet. All the history of game making from the designers and gameplay from the players is stored for further analysis.

\section{Evaluation}
In order to perform a preliminary evaluation of the type and quality of knowledge that GECKA3D can potentially gather, we tested it on 20 university students and staff, who were given the game on tablet and were asked to design a few game scenes over the span of a few hours.Game designers' actions were collected and encoded according to a specific XML format that encodes the semantics associated with such actions (Fig.~\ref{fig:GECKAxml}). 
Specific procedures translate these XML files into pieces of common-sense knowledge to be fed to the SenticNet framework \cite{camsen}. Such pieces of commonsense knowledge were manually evaluated (reasonable VS unreasonable) by 5 annotators, resulting in an accuracy of 85.7\%. They also were asked to try the predefined games on the player mode. 18 out of 20 participants highly appreciate the playability and interestingness of the provided player mode which is crucial for attracting enough players to contribute to the embedded human computing. The other two only care about finishing a task as in traditional GWAPs. More tests are due to verify how such new pieces of common-sense knowledge actually improve the reasoning capabilities of the SenticNet framework on specific tasks such as sentiment analysis.

\begin{table}[h]\conftbl
\begin{tabular}{|x{35pt}|x{28pt}|x{45pt}|x{40pt}|x{28pt}|}\hline
{\bf Item} &{\bf Action} &{\bf Prereq.} &{\bf Outcome} &{\bf Goal}\tn
blender &blend& orange &orange juice &quench thirst\tn
bread &cut &knife &bread slices &--\tn
bread slices &stack &cheese, ham &sandwich &satisfy hunger\tn
coffee beans &hit &pestle &coffee powder &--\tn
coffee maker &fill &coffee powder, boiled water &coffee &--\tn
kettle &fill &water &boiled water &--\tn
chair &hit &hammer &wood pieces &--\tn
can &open &can opener &food &satisfy hunger\tn
towel &cut &scissors &bandage &--\tn
bag &fill &sand &sandbag &flood control\tn
\end{tabular}
\caplab{List of most common POAG triples collected during the pilot testing at Singapore Polytechnic.}{tbl:companies}
\end{table}

\section{Conclusion and Future Work}
In this paper, we have introduced the GECKA3D framework which is used for commonsense knowledge acquisition from game designers through the development of serious games. In fact, this not only provides a platform for the acquisition of commonsense knowledge, but also enables the development of games that transfer back the meaningful lessons to gamers. In the future, we aim to update the gameplay, incentives, and interfaces based on analysis of the data resulting from the game, leading to further understanding of the game itself as well as the nexus between game enjoyment and knowledge generation, e.g., finding out what makes providing information fun or what kinds of information are more `fun' than others. 

\section{Acknowledgments}
This work was supported by Singapore Ministry of Education under research Grants MOE2012-TIF-2-G-015 and MOE2014-TIF-1-G-007.

\bibliographystyle{./styles/aaai}
\bibliography{./refs}

\end{document}